\documentclass[final,5p,times,twocolumn]{elsarticle}

\usepackage{amssymb}
\journal{Astroparticle Physics}

\begin{document}

\begin{frontmatter}



\title{Analysis of Gamma Radiation from a Radon Source: Indications of a Solar Influence}


\author[stan]{P. A. Sturrock\corref{cor}}
\address[stan]{Center for Space Science and Astrophysics, Stanford University, Stanford, CA 94305-4060, USA}
\cortext[cor]{Corresponding author, Tel +1 6507231438; fax +1 6507234840.}
\ead{sturrock@stanford.edu}
\author[GSI]{G. Steinitz}
\address[GSI]{Geological Survey of Israel, Jerusalem, 95501, Israel}
\author[phy]{E. Fischbach}
\address[phy]{Department of Physics, Purdue University, West Lafayette, IN 47907, USA}
\author[412]{D. Javorsek, II}
\address[412]{412th Test Wing, Edwards AFB, CA 93524, USA}
\author[NE,phy]{J. H. Jenkins}
\address[NE]{School of Nuclear Engineering, Purdue University, West Lafayette, IN 47907, USA}

\begin{abstract}
This article presents an analysis of about 29,000 measurements of gamma radiation associated with the decay of radon in a sealed container at the Geological Survey of Israel (GSI) Laboratory in Jerusalem between 28 January 2007 and 10 May 2010. These measurements exhibit strong variations in time of year and time of day, which may be due in part to environmental influences. However, time-series analysis reveals a number of periodicities, including two at approximately 11.2 year$^{-1}$ and 12.5 year$^{-1}$. We have previously found these oscillations in nuclear-decay data acquired at the Brookhaven National Laboratory (BNL) and at the Physikalisch-Technische Bundesanstalt (PTB), and we have suggested that these oscillations are attributable to some form of solar radiation that has its origin in the deep solar interior. A curious property of the GSI data is that the annual oscillation is much stronger in daytime data than in nighttime data, but the opposite is true for all other oscillations. This may be a systematic effect but, if it is not, this property should help narrow the theoretical options for the mechanism responsible for decay-rate variability.
\end{abstract}


\end{frontmatter}

% \linenumbers

\section{Introduction}
\label{sec:Intro}

Radon ($^{222}$Rn) is a radioactive inert gas formed from the disintegration of $^{226}$Ra as part of the $^{238}$U decay chain. It occurs with varying concentrations in the geological environment, and is widely used for tracking temporally varying geological processes \cite{bal91}. The nature of the physical processes driving the temporal patterns observed in $^{222}$Rn time series is not clear, particularly the extent to which environmental parameters, such as temperature, atmospheric pressure, and humidity, can influence the $^{222}$Rn level \cite{cle74,pin97}. To help understand these processes, Steinitz et al. have carried out experiments that track the variation of alpha and gamma radiation arising from a source of $^{222}$Rn in a closed volume of air \cite{ste11}. In this article, we analyze a sequence of 28,733 measurements of gamma radiation acquired in this experiment over the time interval 28 January 2007 to 10 May 2010. The experimental setup is described briefly in Appendix A, and environmental drivers of radon production are discussed in Appendix B.

In Section 2, we present the gross features of the data sequence, which exhibits a strong dependence on time of day and time of year. We also show similar displays of the laboratory temperature and of the solar elevation, finding a stronger correlation of the gamma-detector measurements with the latter than with the former. In Section 3 we present an analysis of the measurements in terms of frequency and time of day, finding that most of the modulation is associated with nighttime measurements. For this reason, we also present separate power-spectrum analyses of the daytime and nighttime measurements. In Section 4, we compare these results with those of our previous time-series analyses of BNL and PTB data \cite{stu10a,jav10,stu10b}, and we comment on a recent study of the decay rate of $^{137}$Cs in an experiment carried out at the Gran Sasso Laboratory \cite{bel12}. Our results are discussed in Section 5.

\section{Overview of the Sequence of Gamma-detector measurements}
\label{sec:Overview}

\begin{figure}[h]
\includegraphics[width=\columnwidth]{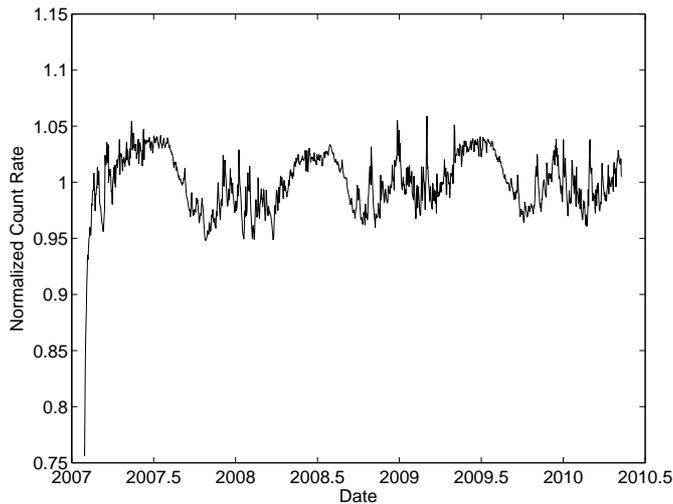}
\caption{Gamma measurements, normalized to mean value unity, as a function of time.  \label{fig:fig1}}
\end{figure}

The mean daily count rate of photons detected by the gamma detector is $2.3610\times{}10^{5}$.  These counts are due primarily to the beta decay of $^{214}$Pb and of $^{214}$Bi, but there are other sources of gamma rays in the $^{222}$Rn decay chain, and one cannot rule out that some counts may be due to the decay of the other $^{222}$Rn daughters. 
Figure \ref{fig:fig1} shows a plot of the daily count, normalized to mean value unity, as a function of time. Since the early values are anomalous, we exclude them from further consideration.

We next remove fluctuations on the time scale of one day by forming 21-point running means. Figure 2 shows the percentage departure from the mean value (unity).  The dominant feature is an annual modulation, with a peak at mid-summer. However, it is notable that there is also a minor peak near mid-winter.

\begin{figure}[h]
\includegraphics[width=\columnwidth]{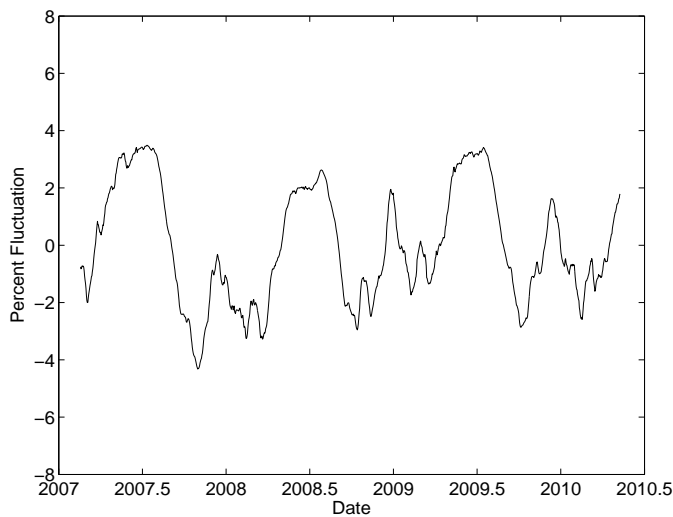}
\caption{Fluctuation (in percent) of the 21-point running mean of the (abbreviated) gamma measurements.   \label{fig:fig2}}
\end{figure}

Due to the experimental setup and location, it is reasonable to suspect that an annual modulation may have an environmental origin. We therefore show comparable plots derived from the absolute temperature (mean value 291.87 K) in Figure 3; pressure (mean value 923.78 milli-bar) in Figure 4; and line voltage (mean value 13.403 Volts) in Figure 5.

\begin{figure}[h]
\includegraphics[width=\columnwidth]{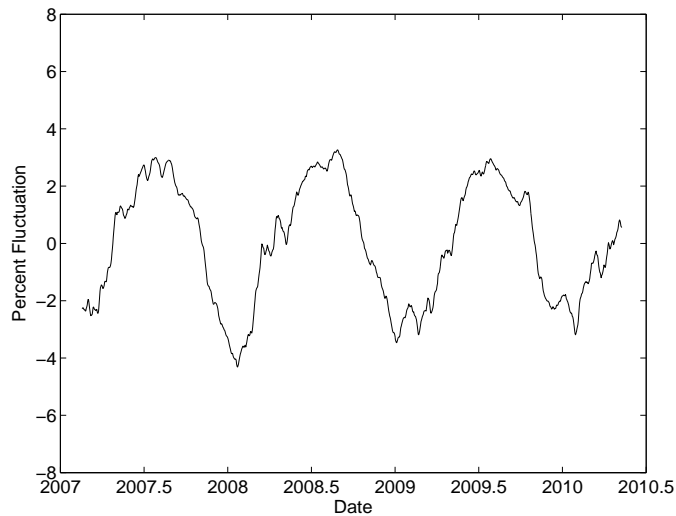}
\caption{Fluctuation (in percent) of the 21-point running mean of the absolute temperature.  \label{fig:fig3}}
\end{figure} 

\begin{figure}[h]
\includegraphics[width=\columnwidth]{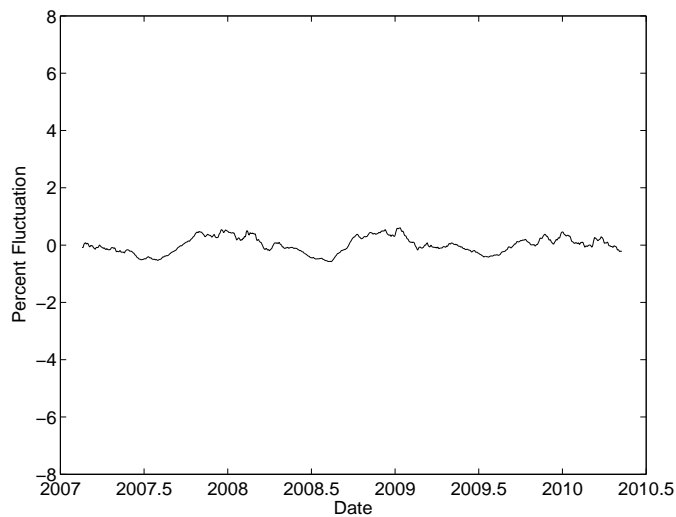}
\caption{Fluctuation (in percent) of the 21-point running mean of the pressure.   \label{fig:fig4}}
\end{figure} 

\begin{figure}[h]
\includegraphics[width=\columnwidth]{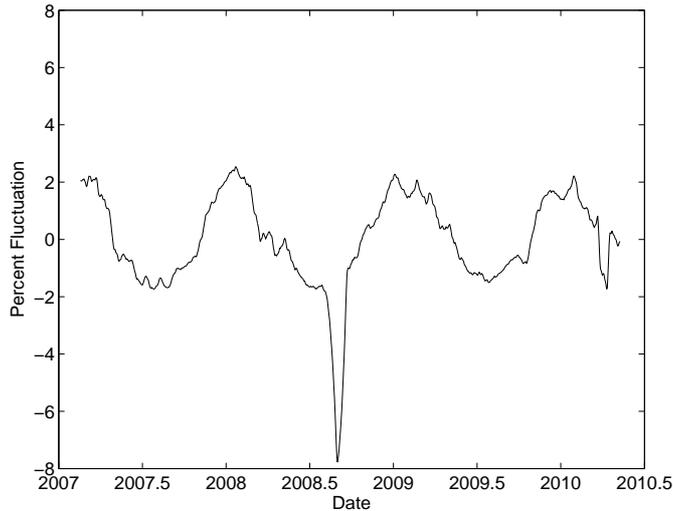}
\caption{Fluctuation (in percent) of the 21-point running mean of the voltage.  \label{fig:fig5}}
\end{figure} 

Pressure shows little annual variation, but temperature and voltage both show annual modulations with amplitudes and phases comparable with that of the gamma-detector measurements, except that the voltage curve is in anti-phase. Table \ref{tab:tbl1} lists the amplitude of the sine-wave component of each of these wave-forms, together with the phase-of-year of the maximum and the phase-of-year of the minimum for each sine-wave fit. We see that the amplitudes of the temperature and voltage variations are comparable with that of the gamma-detector measurements. However, even allowing for an inversion of the voltage curve, the phases differ by over one month. Such a large phase shift would be difficult to understand if the annual variation of the gamma-detector measurements were caused by the annual variations of either the temperature or the voltage.

\begin{table}[h]
\label{tab:tbl1}
\caption{Amplitudes and Phases of Gamma, Temperature, Pressure, and Voltage.}
\footnotesize
\begin{tabular}{lccc}
\hline 
 & Amplitude & Phase of Maximum & Phase of Minimum \\
\hline Gamma & 0.0208 & 0.4664 & 0.9664 \\
Temperature&0.0287&0.5660&0.0660 \\
Pressure&0.0036&0.9895&0.4895 \\
Voltage&0.0196&0.0770&0.5770 \\
\hline 
\end{tabular} 
\end{table}
\normalsize

These phase relationships are also evident in Figure 6 and 7. Figure 6 combines the waveforms of the gamma-detector measurements and of the temperature. Figure 7 combines the waveform of the gamma-detector measurements and the inverted waveform of the voltage. It is notable that the temperature and voltage curves lag the gamma curve.

\begin{figure}[h]
\includegraphics[width=\columnwidth]{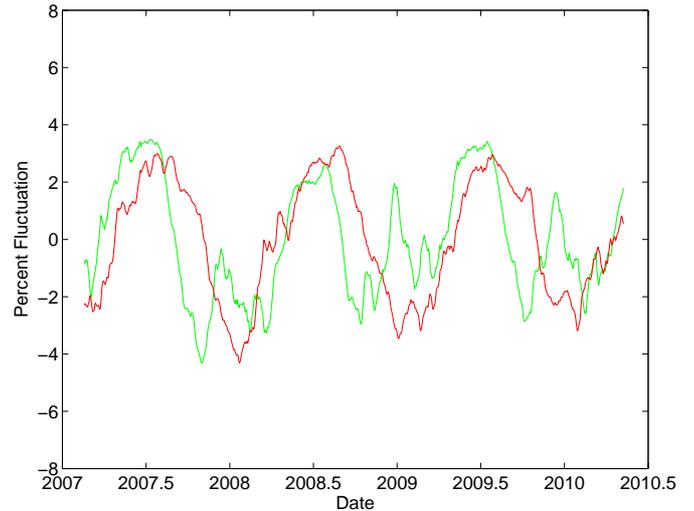}
\caption{Fluctuation (in percent) of the 21-point running mean of the absolute temperature (red) and of the gamma measurements (green).  \label{fig:fig6}}
\end{figure} 

\begin{figure}[h]
\includegraphics[width=\columnwidth]{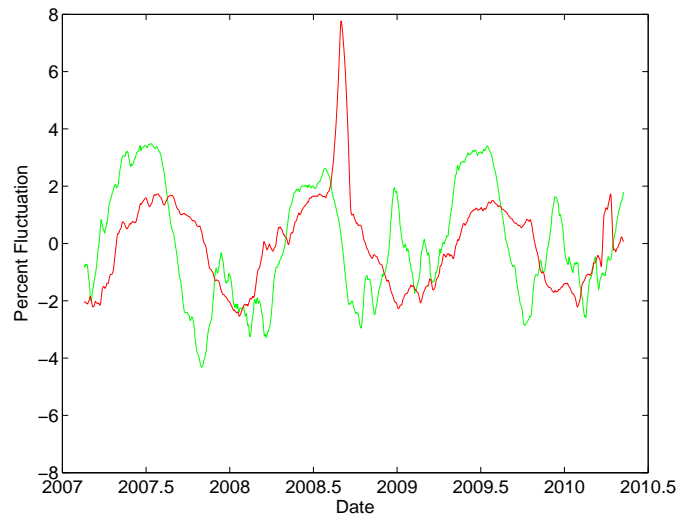}
\caption{Fluctuation (in percent) of the 21-point running mean of the voltage (inverted, red) and of the gamma measurements (green).   \label{fig:fig7}}
\end{figure}

The variation of the gamma-detector measurements as a function of both date and time of day is shown in Figure 8. We see that there is a strong dependence on time of day, as well as on time of year. The variation of temperature as a function of date and time of day is shown in Figure 9. The corresponding plot for solar elevation is shown in Figure 10. The solar elevation has its maximum value close to mid-year and mid-day, as expected. The peaks in the gamma-detector measurements occur close to mid-year, but a little later than mid-day. The peaks in the temperature measurements occur slightly later than mid-year and slightly later than mid-day.

\begin{figure}[h]
\includegraphics[width=\columnwidth]{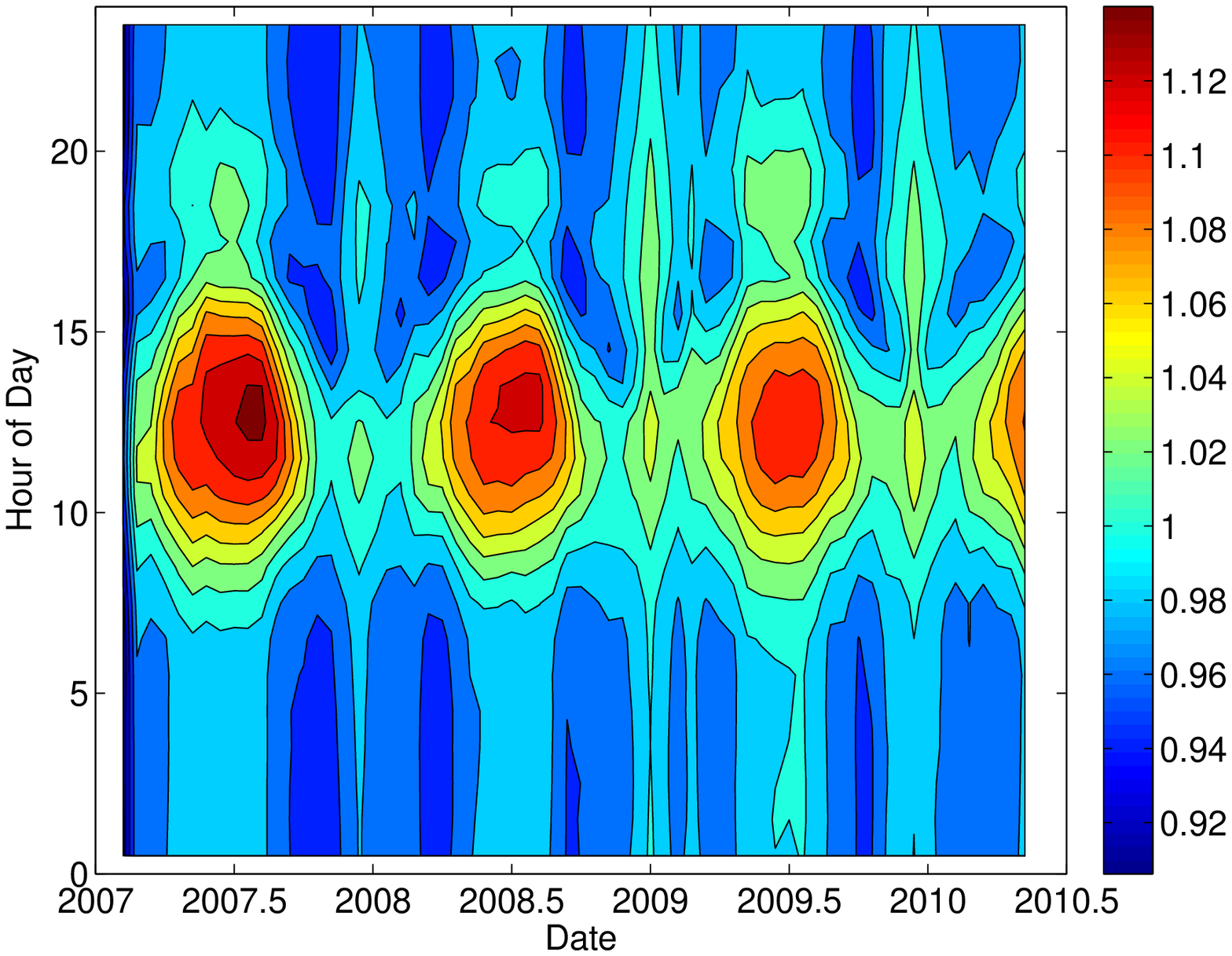}
\caption{Gamma measurements as a function of date and time of day. The color-bar gives the power, $S$, of the observed signal. \label{fig:fig8}}
\end{figure} 

\begin{figure}[h]
\includegraphics[width=\columnwidth]{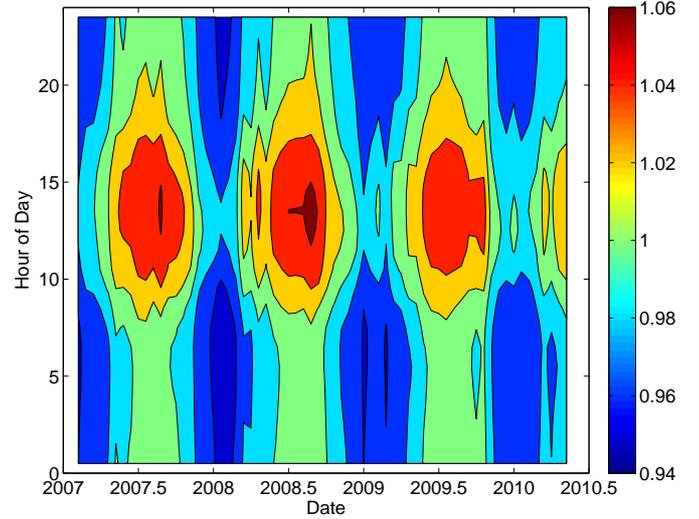}
\caption{Temperature as a function of date and time of day. The color-bar gives the power, $S$, of the observed signal.   \label{fig:fig9}}
\end{figure} 

\begin{figure}[h]
\includegraphics[width=\columnwidth]{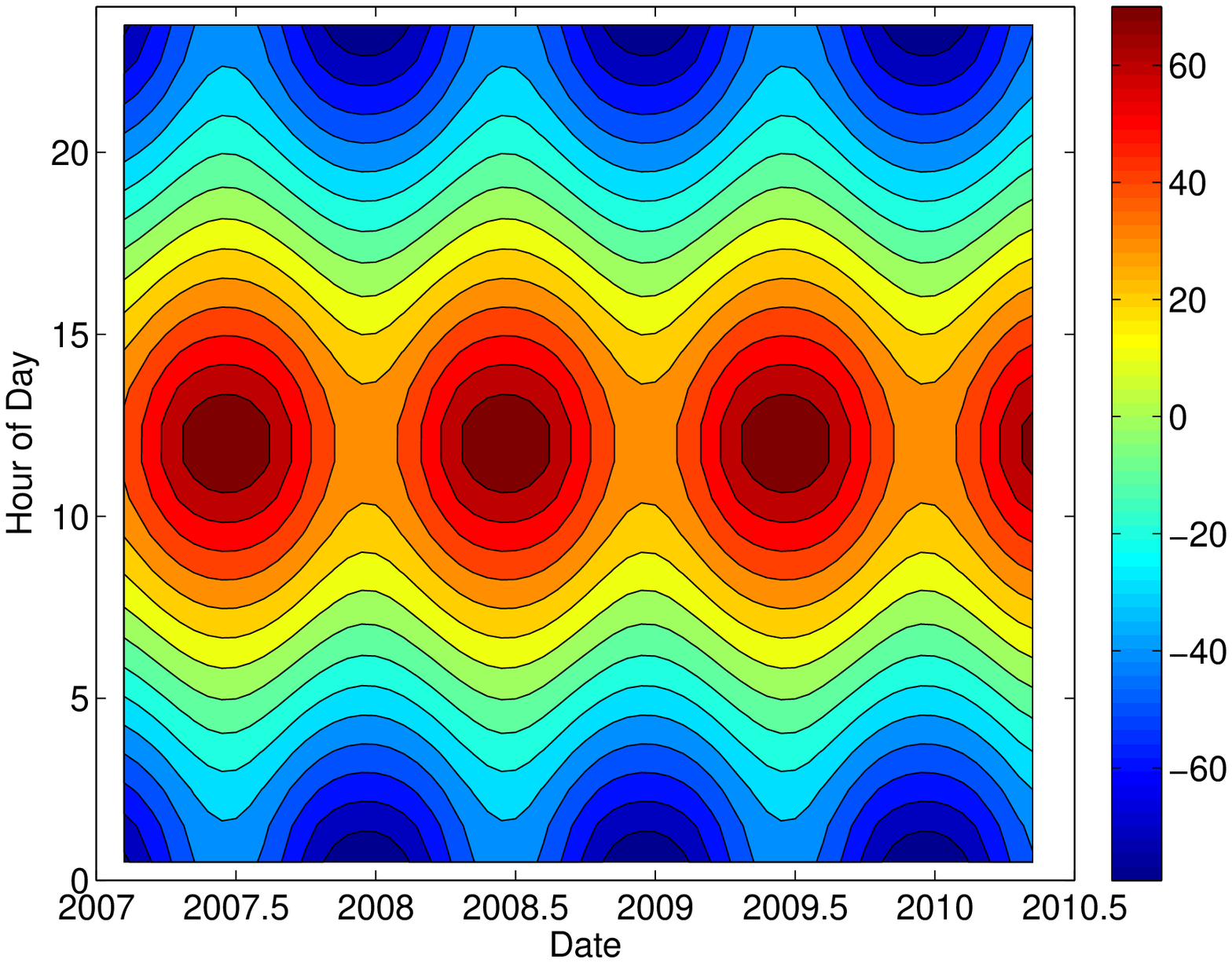}
\caption{Solar elevation as a function of date and time of day. The color-bar gives the power, $S$, of the observed signal.  \label{fig:fig10}}
\end{figure} 

These results suggest that the variation in the gamma-detector measurements is related to environmental conditions, but not in exactly the same way as is temperature.

\section{Power Spectrum Analyses}
\label{sec:PSA}

We have carried out a power-spectrum analysis of the gamma-detector measurements when broken down by hour of day. For the frequency band 0-5 year$^{-1}$, the result is shown in Figure 11. The strongest feature corresponds to an annual oscillation centered on mid-day. The next strongest feature (frequency 2 year$^{-1}$) is a semi-annual oscillation. This is mainly a night-time feature, but it has a peak at 17 hours.

\begin{figure}[h]
\includegraphics[width=\columnwidth]{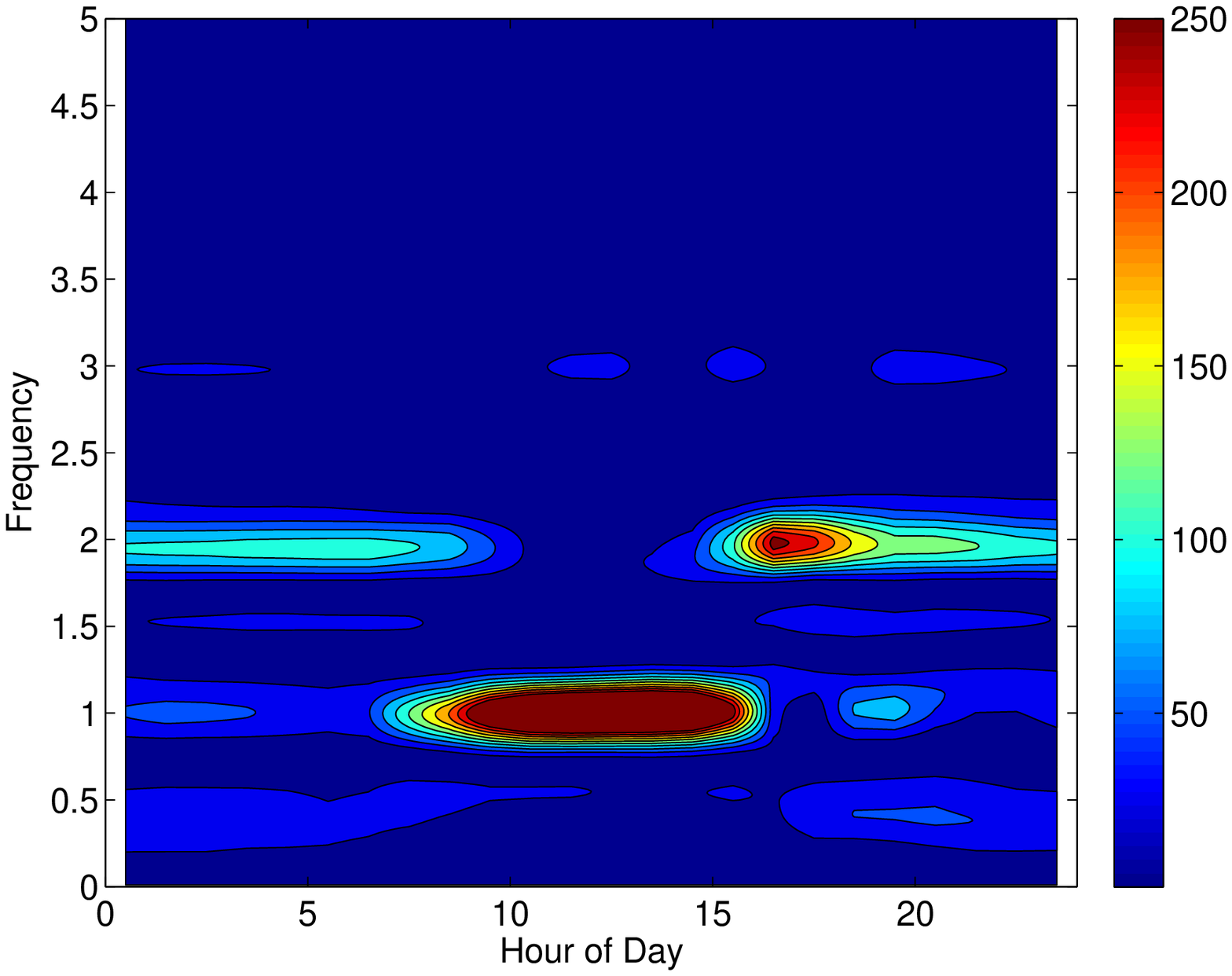}
\caption{Frequency-Hour-Of-Day display for the frequency range 0-5 year$^{-1}$. The color-bar gives the power, $S$, of the observed signal.   \label{fig:fig11}}
\end{figure} 

Figure 12 covers the frequency band 5-10 year$^{-1}$. The oscillations in this band all occur at nighttime. The strongest feature is at frequency 8.9 year$^{-1}$.

\begin{figure}[h]
\includegraphics[width=\columnwidth]{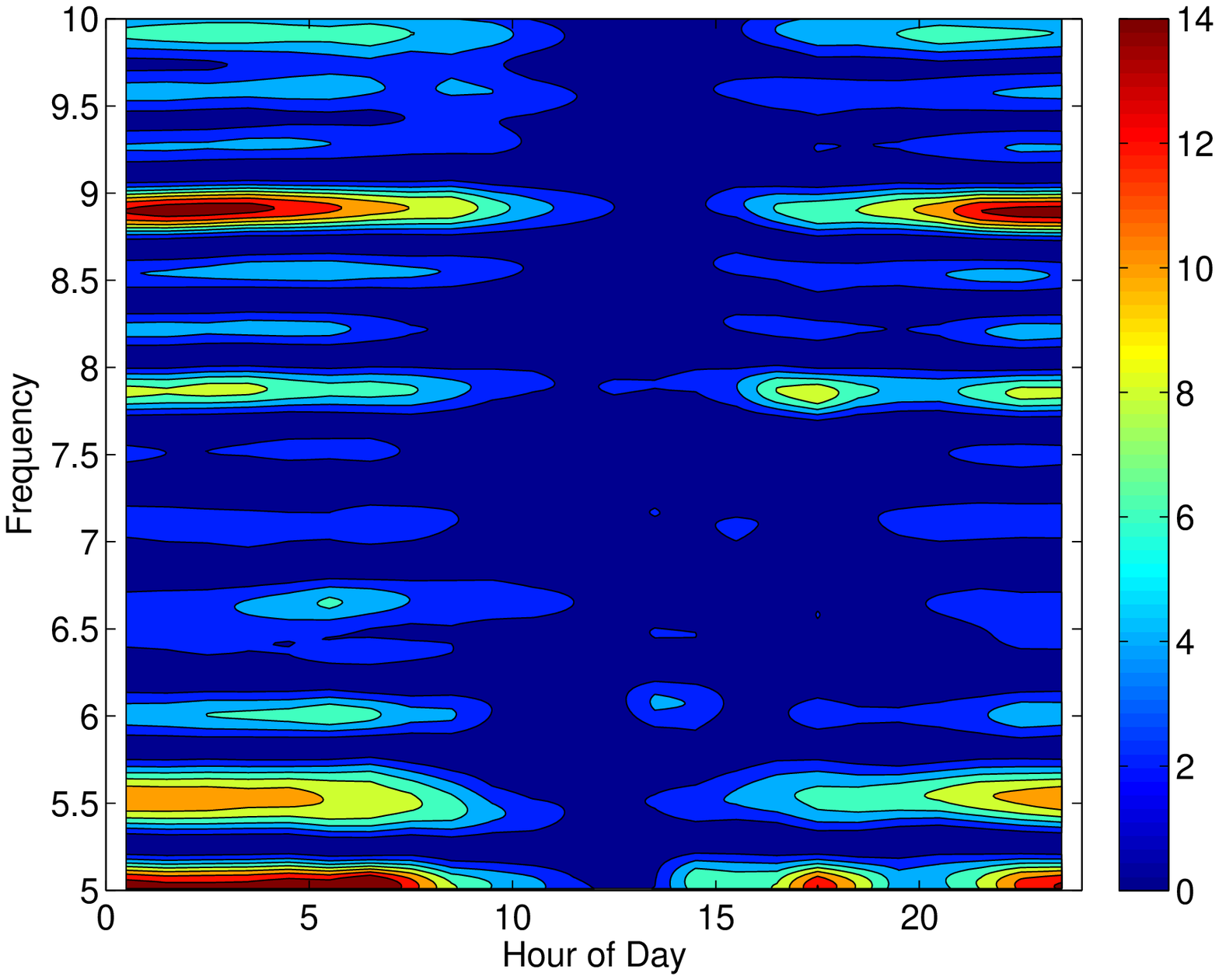}
\caption{Frequency-Hour-Of-Day display for the frequency range 5-10 year$^{-1}$. The color-bar gives the power, $S$, of the observed signal.  \label{fig:fig12}}
\end{figure} 

Figure 13 covers the frequency band 10-15 year$^{-1}$. Here also, the oscillations all occur at nighttime. The strongest feature is at frequency 12.5 year$^{-1}$, but there is also a feature at 11.3 year$^{-1}$ and a lesser feature at 11.9 year$^{-1}$.

\begin{figure}[h]
\includegraphics[width=\columnwidth]{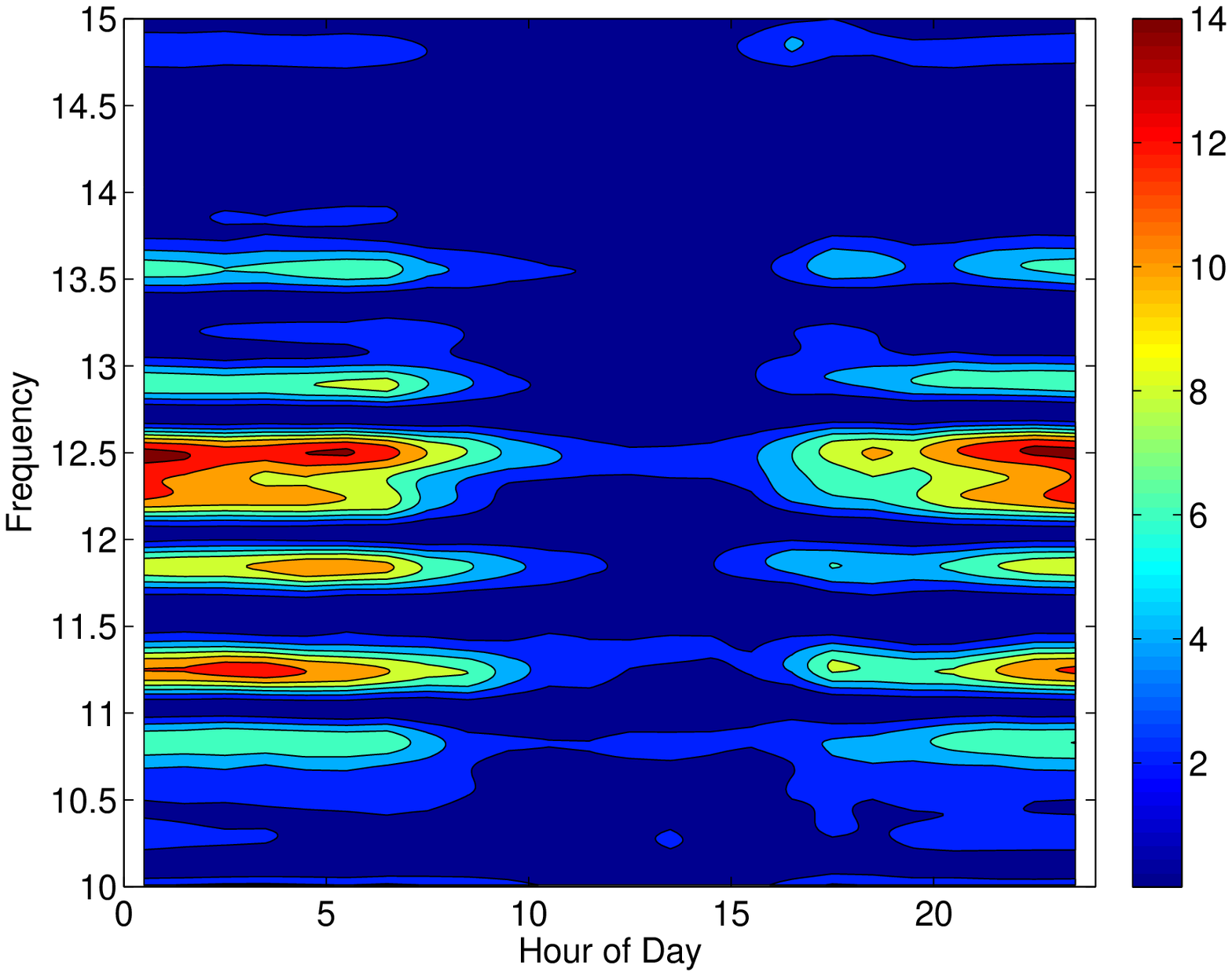}
\caption{Frequency-Hour-Of-Day display for the frequency range 10-15 year$^{-1}$. The color-bar gives the power, $S$, of the observed signal.   \label{fig:fig13}}
\end{figure} 

It is clear from the preceding results that the power-spectrum properties of the gamma-detector measurements will be quite different for daytime measurements and nighttime measurements. We therefore examine separately data acquired over the time interval 6 to 18 hours (daytime), and from 18 hours to 6 hours (nighttime).

We show in Figure 14 the power spectra derived from daytime and nighttime measurements over the frequency range 0-5 year$^{-1}$. We see that there is a very strong peak at 1 year$^{-1}$ in the daytime data, and a strong peak at 2 year$^{-1}$ in the nighttime data, as we would expect from Figure 11.

\begin{figure}[h]
\includegraphics[width=\columnwidth]{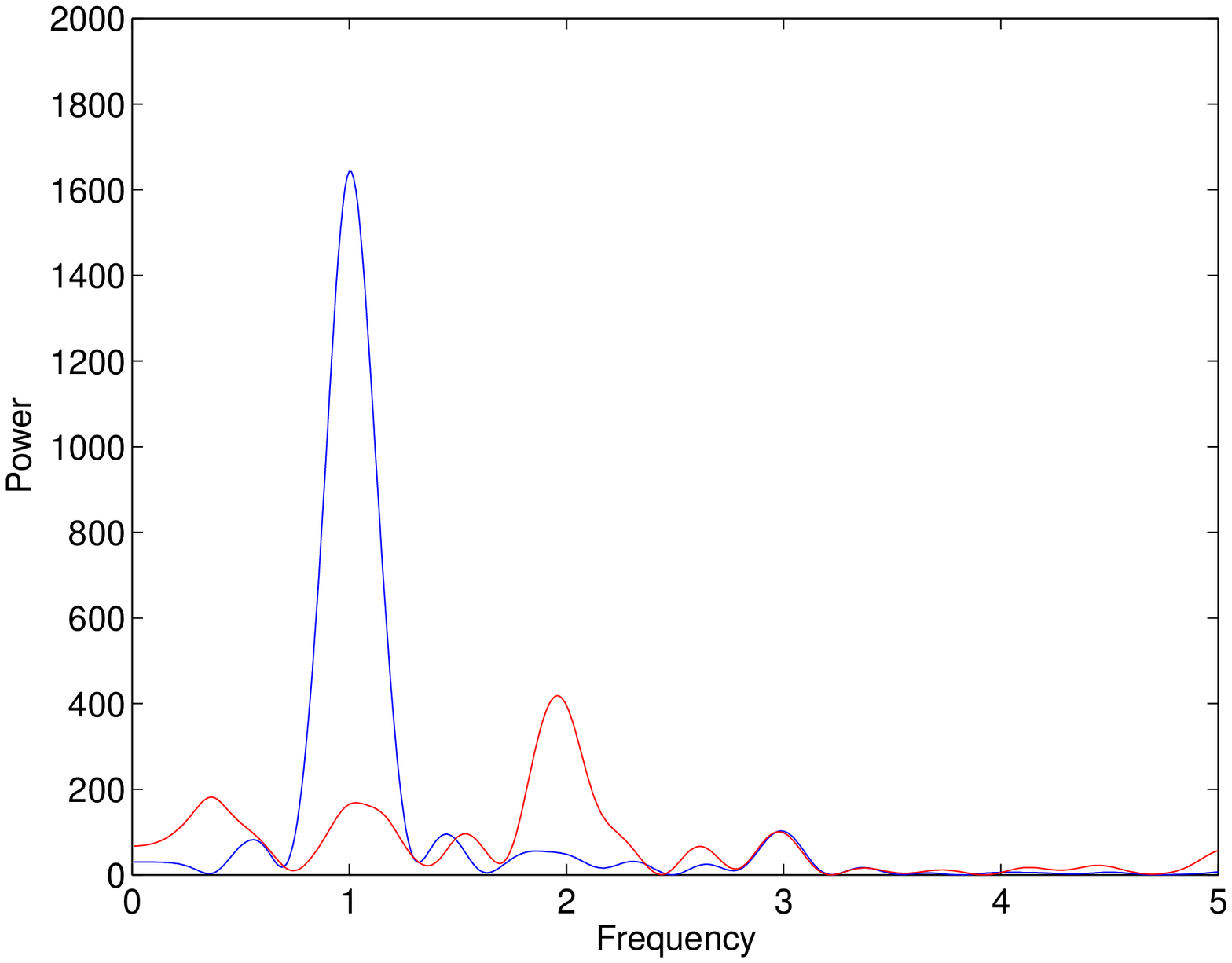}
\caption{Power spectra over the frequency range 0-5 year$^{-1}$. Power from daytime data shown negative, and power from nighttime data shown positive.  \label{fig:fig14}}
\end{figure} 

We see from Figure 15 that, for the frequency range 5-20 year$^{-1}$, the nighttime data lead to significantly stronger oscillations than do the daytime data. However, the fine structure in these power spectra seems unrealistic, since the duration of the time series is only about 3 years. For this reason, we have formed a 41-point running mean of the power spectrum, resulting in the plot shown in Figure 16. The top 20 peaks in the nighttime power spectrum are listed in Table 2. We see from this figure and from this table that there are two notable peaks in the rotational search band 10-15 year$^{-1}$, one at 12.39 year$^{-1}$, with power $S$=49.63, and the other at 11.27 year$^{-1}$, with power $S$=29.39. (In power-spectrum analysis, the probability of the null hypothesis, that the peak is due to normally distributed random noise, is given by $e^{-S}$, where $S$ is the power value of the frequency peak \cite{sca82}.)

\begin{table}
\label{tbl:tbl2}
\caption{Top 20 peaks in the 41-point running mean of the power spectrum formed from nighttime data.}
\footnotesize
\begin{tabular}{lcc}
\hline
Frequency (year$^{-1}$)&Power&Order\\
\hline
1.99&286.22&1\\
0.37&137.00&2\\
1.06&130.99&3\\
2.89&57.21&4\\
12.39&49.63&5\\
5.00&35.65&6\\
5.54&30.27&7\\
11.27&29.39&8\\
8.90&29.38&9\\
17.54&22.18&10\\
10.76&20.12&11\\
7.96&17.87&12\\
7.80&17.28&13\\
4.32&16.46&14\\
18.53&16.07&15\\
12.98&15.52&16\\
13.62&15.34&17\\
9.80&15.01&18\\
18.15&15.00&19\\
19.68&13.41&20\\
\hline
\end{tabular}
\normalsize
\end{table}

\begin{figure}[h]
\includegraphics[width=\columnwidth]{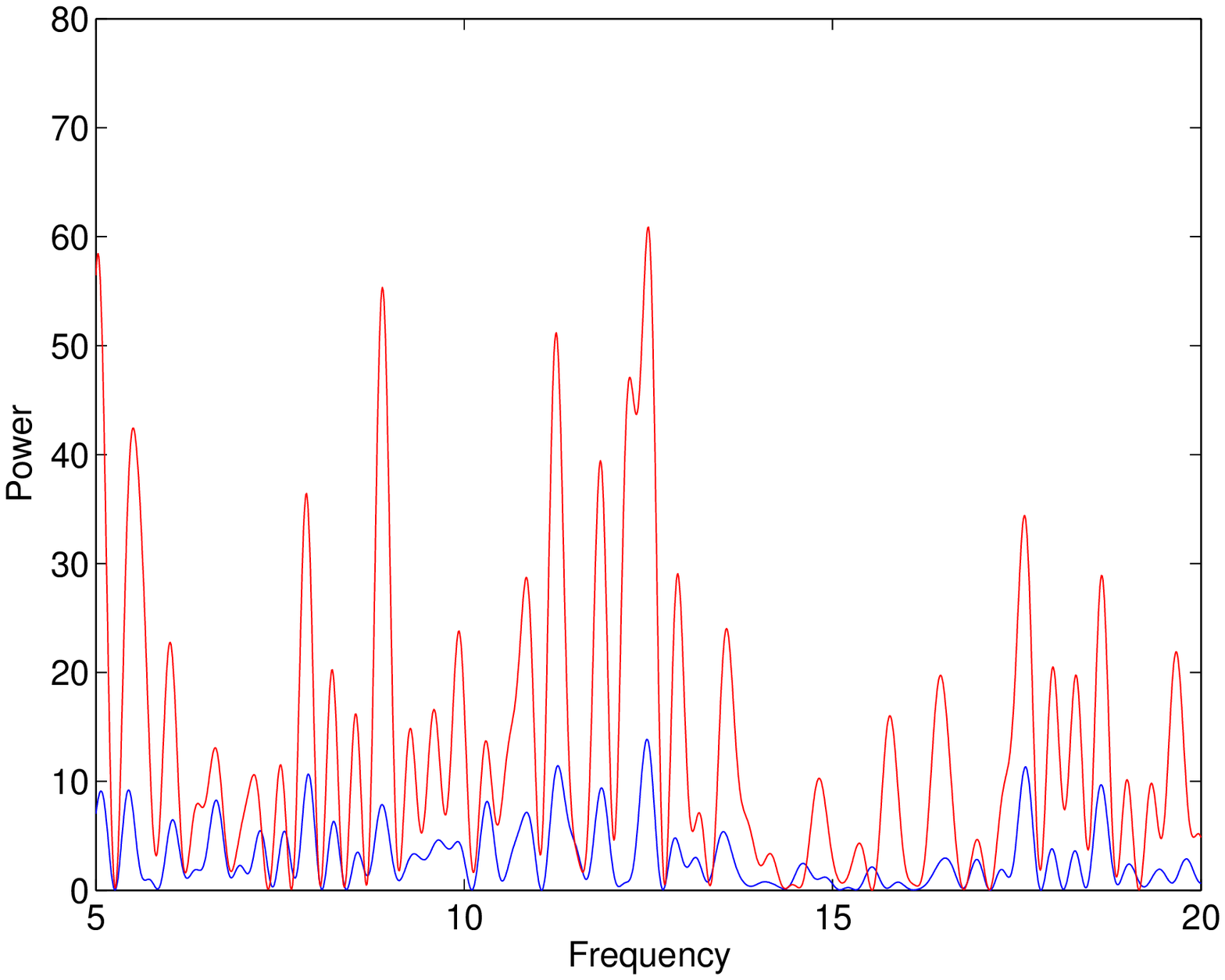}
\caption{Power spectra over the frequency range 5-20 year$^{-1}$. Power from daytime data shown in blue, and power from nighttime data shown in red.  \label{fig:fig15}}
\end{figure} 

\begin{figure}[h]
\includegraphics[width=\columnwidth]{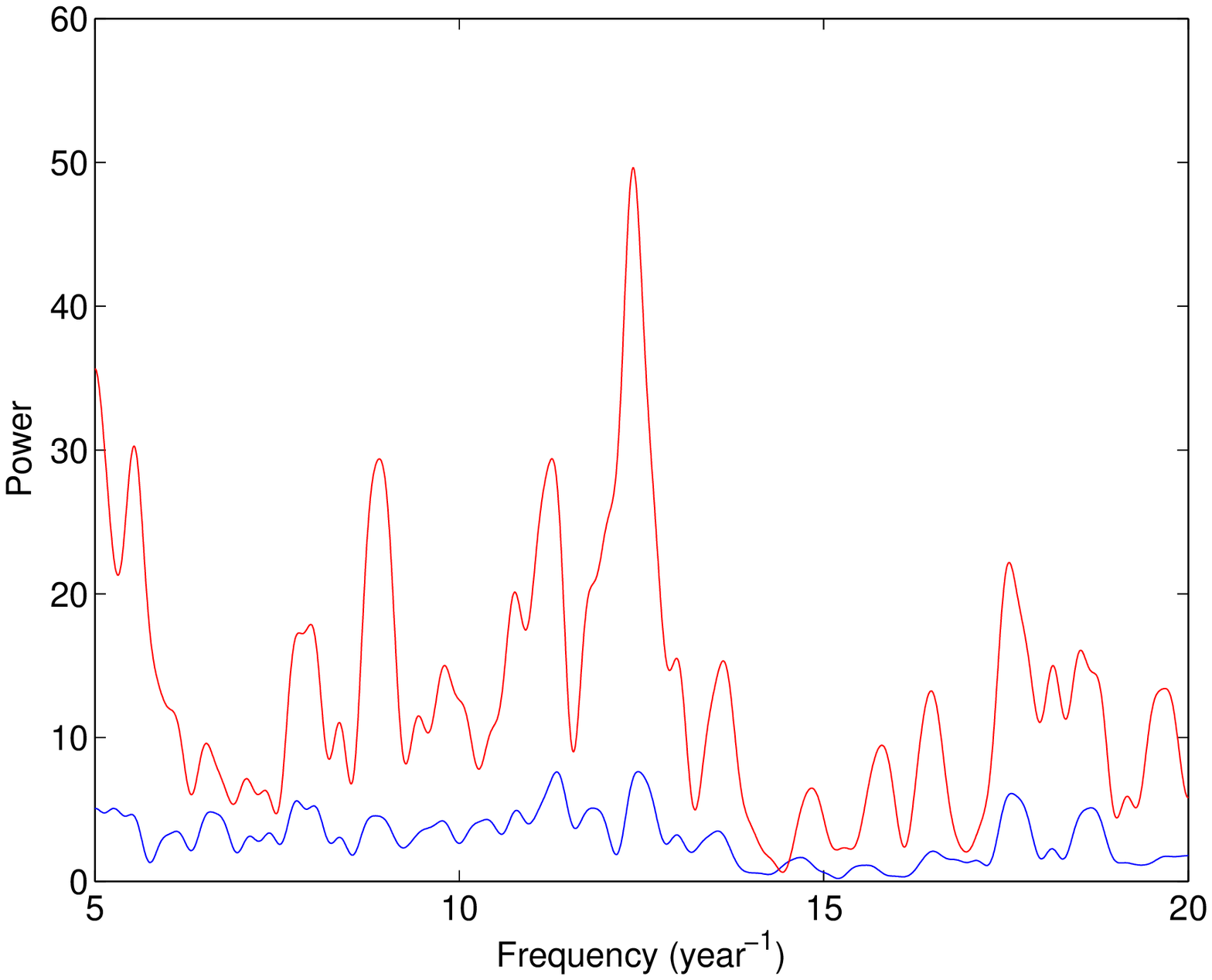}
\caption{41-point running mean of the power spectra over the frequency range 5-20 year$^{-1}$. Power from daytime data shown in blue, and power from nighttime data shown in red.   \label{fig:fig16}}
\end{figure} 

It is interesting, and possibly significant, that the frequencies of the 8th and 7th peaks (11.27 year$^{-1}$ and 5.54 year$^{-1}$) are in close to a 2:1 ratio.

In previous analyses of decay data \cite{stu10a,jav10,stu10b}, we have evaluated the significance of a peak in the rotational search band by using the shuffle test \cite{bah91}. We have therefore carried out 1,000 shuffle simulations of the nighttime data, with the result shown in histogram form in Figure 17. The largest value of the power generated in this way is 9.35, far below the actual value of 49.63. Hence there is no possibility that the oscillations are the result of independent random fluctuations in the measurements.

\begin{figure}[h]
\includegraphics[width=\columnwidth]{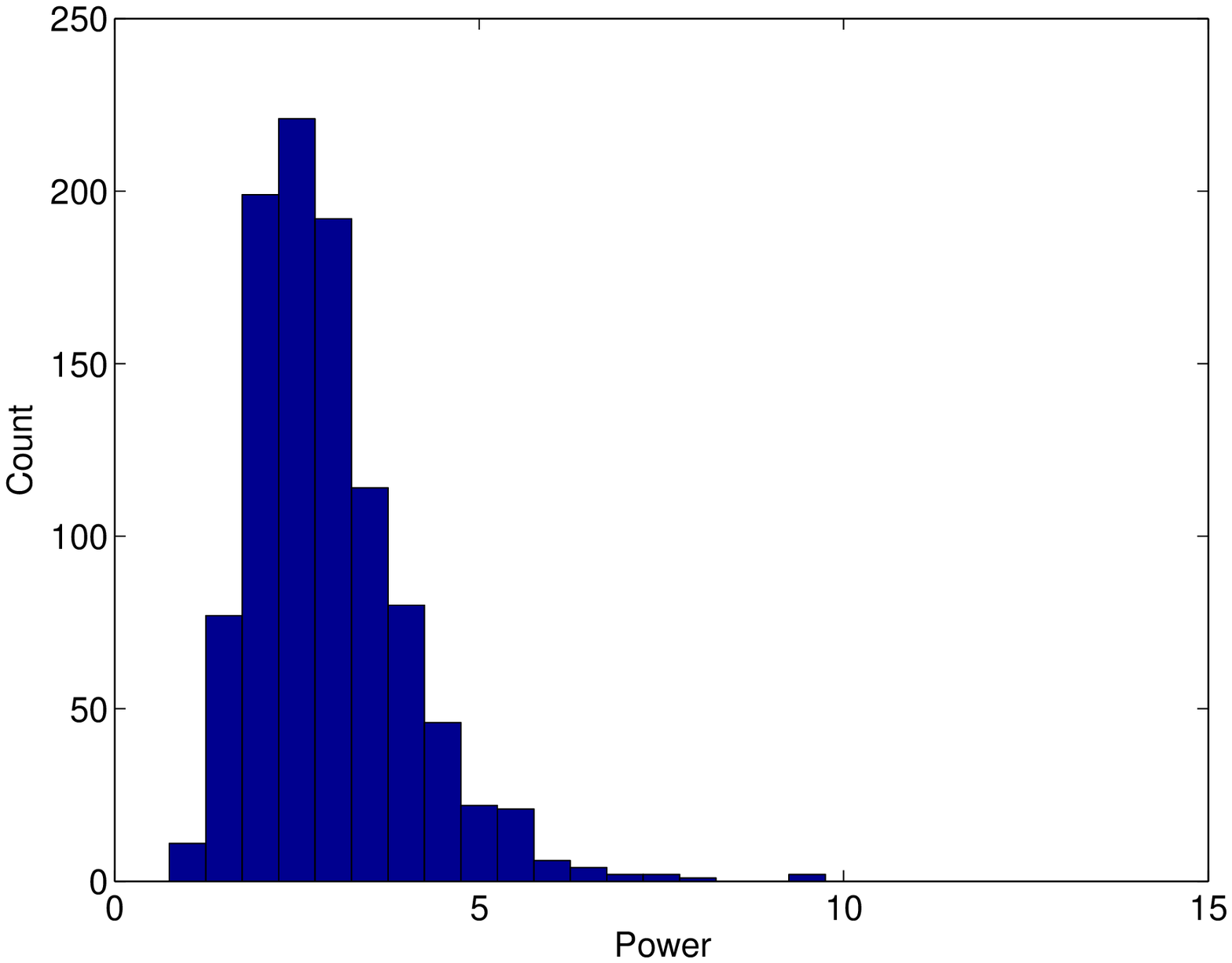}
\caption{Histogram formed from 1,000 shuffle simulations of the 41-point running mean of the power spectrum formed from nighttime data. None has power in the rotational band as large as 49.63. The largest value in the simulated data is 9.35.  \label{fig:fig17}}
\end{figure} 

It is clear from Figures 15 and 16 that the nighttime data contains oscillations that are not present, or only weakly present, in the daytime data, and that the difference between the daytime and nighttime power spectra is statistically significant.

\section{Comparison with BNL and PTB Data}
\label{sec:BNL+PTB}

It is interesting to compare the results of our analysis of the GSI data with the results of similar analyses of the BNL (Brookhaven National Laboratory) data \cite{alb86} and PTB (Physikalisch-Technische Bundenanstalt) \cite{sie98} data. We have previously shown the results of simple power-spectrum analyses of these two datasets \cite{stu10a,jav10,stu10b}. However, to facilitate visual comparison with our present analysis of GSI data, we now show instead the results of time-frequency analyses.

We show in Figures 18 and 19 time-frequency displays of the BNL measurements of the decay rates of $^{36}$Cl and $^{32}$Si, respectively, for the frequency band 10-15 year$^{-1}$. We see that both of these figures exhibit features in the two bands 11.0 to 11.4 year$^{-1}$ and 12.6 to 12.9 year$^{-1}$. 

\begin{figure}[h]
\includegraphics[width=\columnwidth]{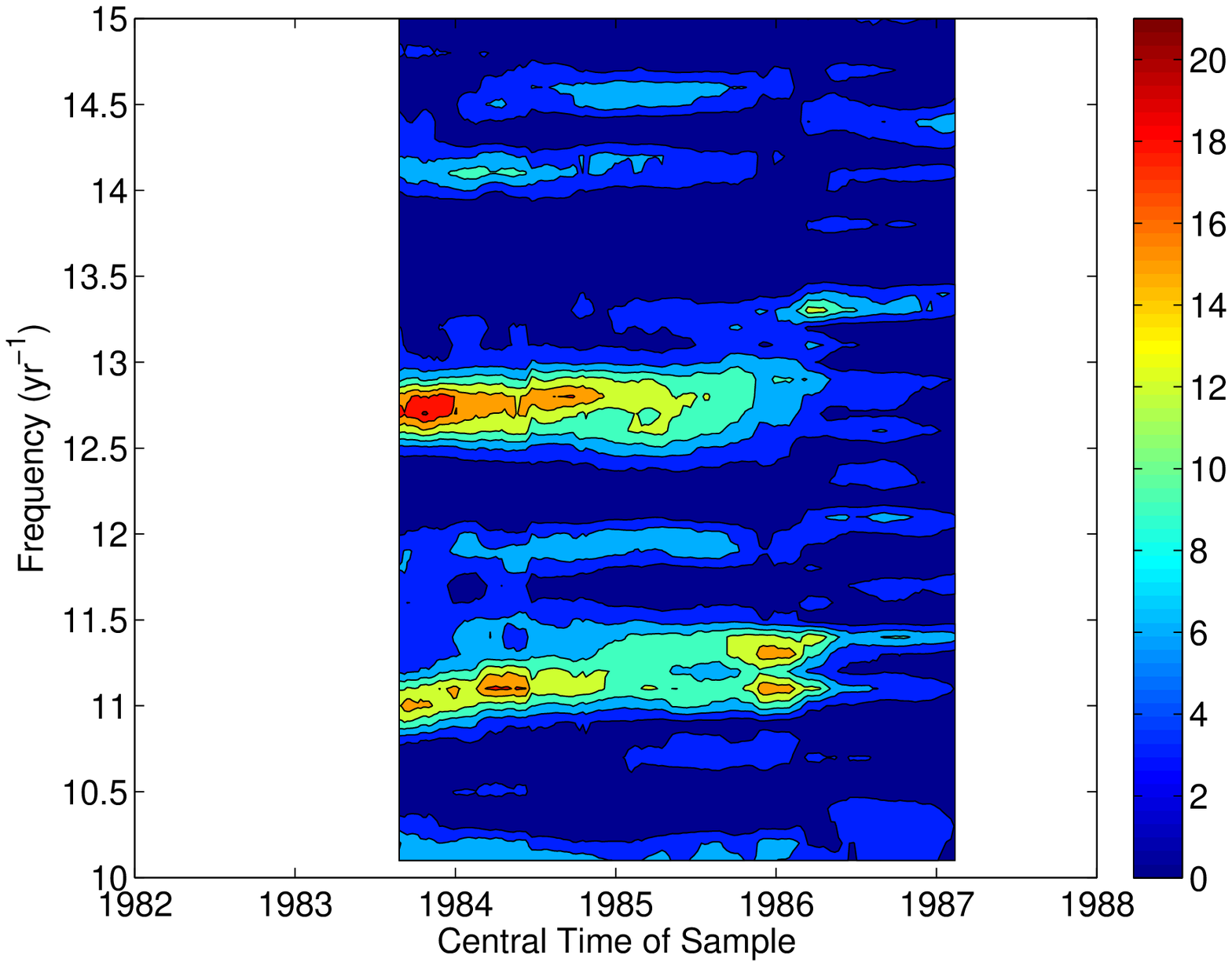}
\caption{Frequency-time display of the $^{36}$Cl decay measurements made by the BNL (Brookhaven National Laboratory) experiment. The color-bar gives the power, $S$, of the observed signal.   \label{fig:fig18}}
\end{figure} 

\begin{figure}[h]
\includegraphics[width=\columnwidth]{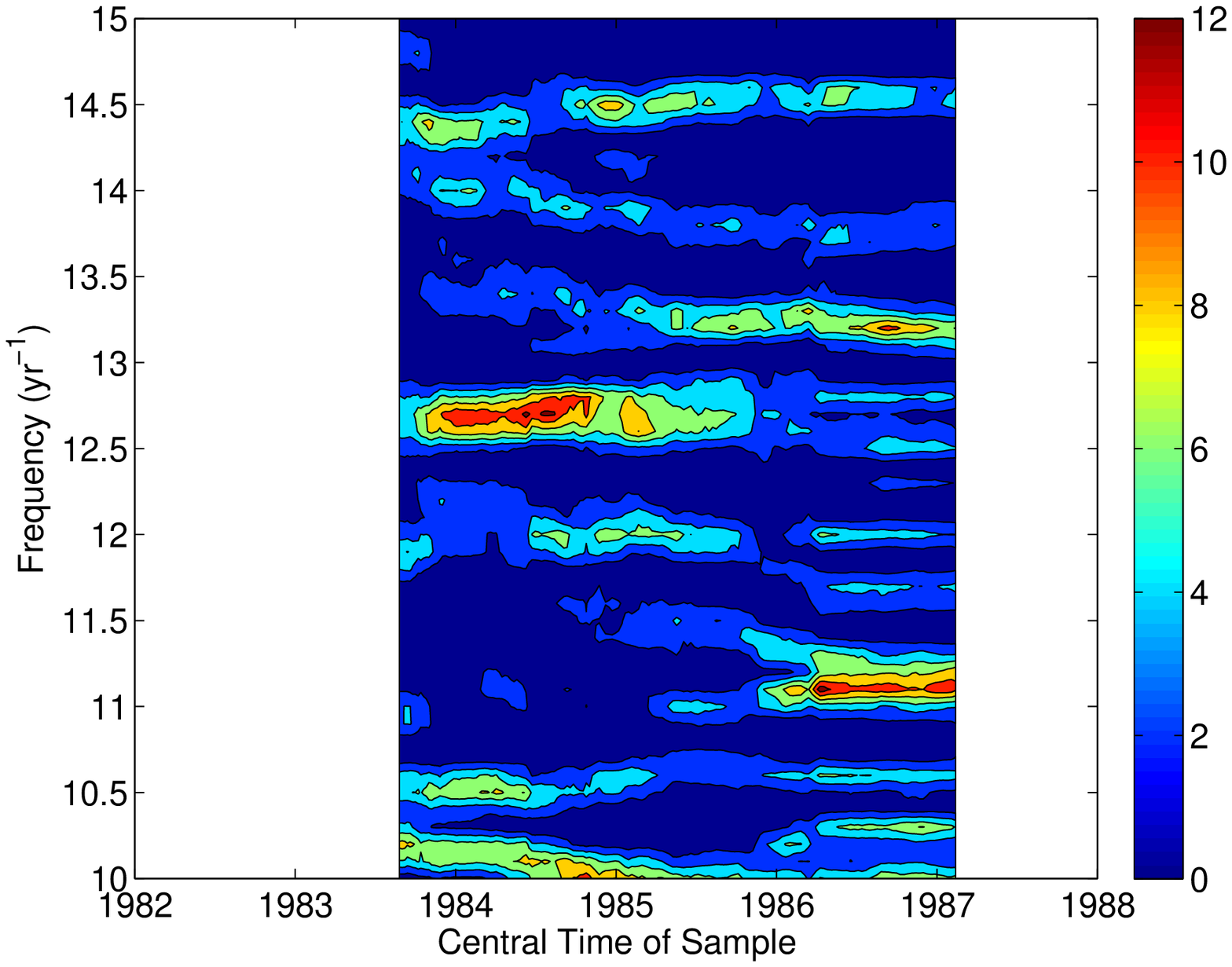}
\caption{Frequency-time display of the $^{32}$Si decay measurements made by the BNL (Brookhaven National Laboratory) experiment.  The color-bar gives the power, $S$, of the observed signal.  \label{fig:fig19}}
\end{figure} 

We show in Figure 20 a time-frequency display of the PTB $^{226}$Ra data. We see two features in the frequency range 11.0 to 11.3 year$^{-1}$, and another feature in the range 12.3 to 12.5 year$^{-1}$.

\begin{figure}[h]
\includegraphics[width=\columnwidth]{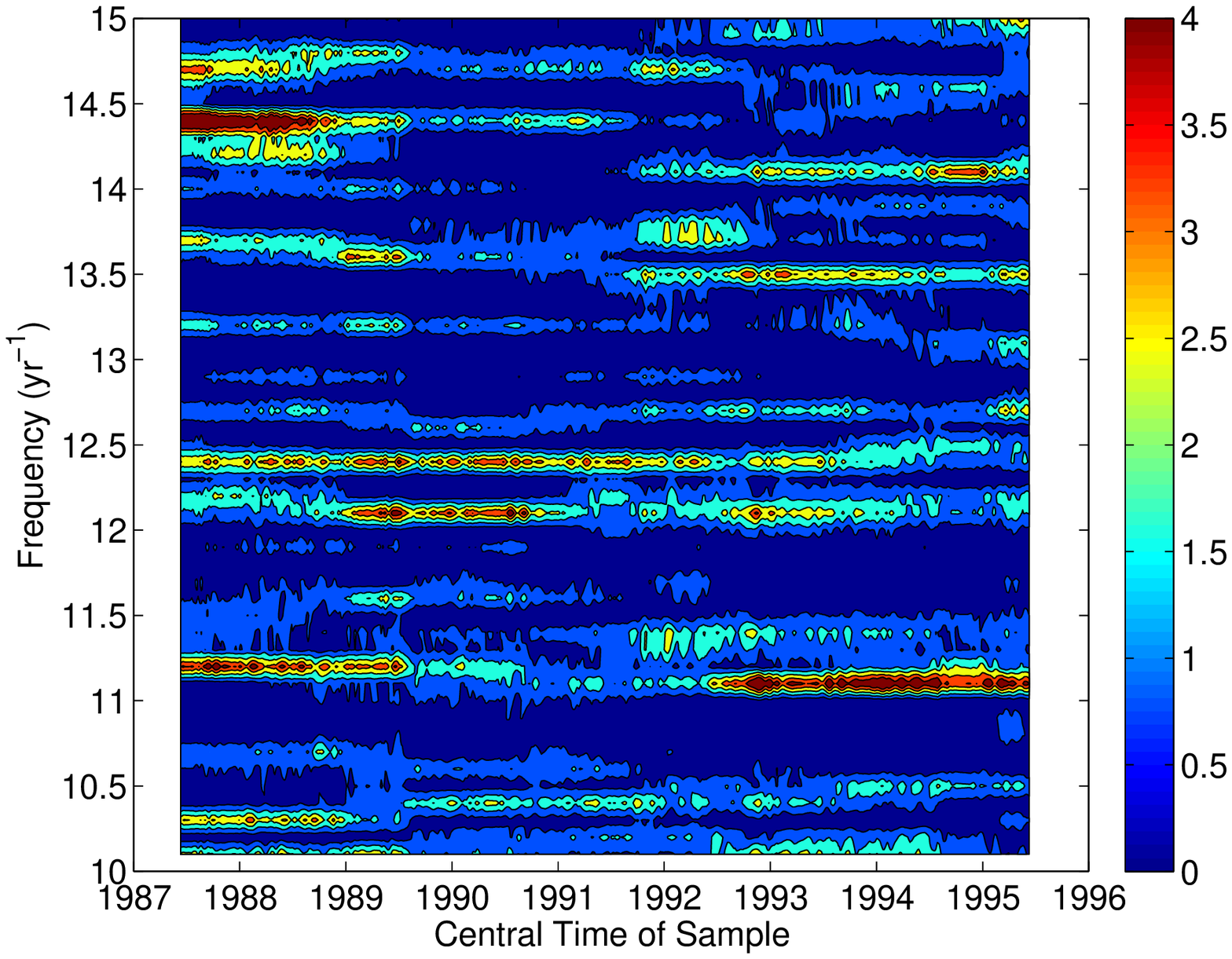}
\caption{Frequency-time display of the $^{226}$Ra decay measurements made by the PTB (Physikalisch-Technische Bundesanstalt) experiment. The color-bar gives the power, $S$, of the observed signal.  \label{fig:fig20}}
\end{figure} 

For comparison, we see in Figure 13 (formed form GSI data) a feature in the band 11.1 to 11.4 year$^{-1}$, and a feature in the band 12.2 to 12.6 year$^{-1}$. We see that all four datasets lead to oscillations in the two bands 11.0 to 11.4 year$^{-1}$ and 12.1 to 12.9 year$^{-1}$. 

It should be noted that the BNL and PTB oscillations are intermittent in amplitude, and drift somewhat in frequency, but we nevertheless see broad agreement among oscillation patterns in all four displays.

Bellotti et al. have recently reported the result of measurements of the activity of a $^{137}$Cs source, as determined by an experiment installed deep underground in the Gran Sasso laboratory \cite{bel12}, finding no evidence of variability in that dataset.  This result is not inconsistent with our studies to date, which have shown that (a) not all nuclides exhibit variability; and (b) among nuclides that do exhibit variability, the patterns of variability are not all the same.

\section{Discussion}
\label{sec:Disc}

The patterns we see in the GSI radon measurements are so striking as to be quite incompatible with the textbook picture that nuclear decay rates are all constant. The first question to be addressed is the extent to which the variations have an environmental or experimental origin. Figures 6, 7, 8 and 9 show that the gross variations of the gamma-detector measurements are quite similar to those of the ambient temperature. However, the annual waveform of the gamma-detector measurements leads  rather than lags that of the ambient temperature. This makes it unlikely that the annual variation of the decay rates is due completely to the variation in the ambient temperature. Figure 7 shows that the annual waveform of the gamma-detector measurements also leads that of the voltage. Hence it is also unlikely that the annual variation of the decay rates is completely due to the voltage variation.

On the other hand, comparison of Figures 9 and 10 shows that the annual variation of the gamma-detector measurements lags that of the solar elevation, which is compatible with the possibility that the annual variation of decay rates may be due to a direct solar influence rather than a secondary environmental influence. Note, however, that Figure 10 represents not only the maximum elevation as seen at the site of the experiment, but also the maximum elevation as it would be seen at the diametrically opposite position on Earth.

The frequency-hour-of-day displays in Figures 11, 12, and 13 show that the annual oscillation is primarily a daytime effect, but all other oscillations are primarily nighttime effects. It may turn out that this curious property of the time series is due to some experimental or environmental effect, but we have so far failed to identify such an effect. If this pattern proves not to be attributable to an experimental or environmental effect, indicating that it is an intrinsic property of the decay process, we would find it difficult to avoid the conclusion that the relevant decay process is anisotropic. There would then be a clear need for a dedicated experiment that could distinguish between isotropy and anisotropy.

We have noted in our earlier analyses \cite{stu10a,jav10,stu10b} of BNL and PTB data \cite{alb86,sie98} that the above oscillation frequencies are too low to be attributable to rotation in either the convection zone or the radiative zone, leading to the suggestion that the solar core may rotate more slowly than the outer layers. This conjecture leads in turn to the suggestion that there may be an inner tachocline (a region involving a sharp gradient in the rotation rate) that separates the core from the radiative zone, analogous to the well known outer tachocline that separates the radiative zone from the convection zone. An equatorial section of the outer tachocline extends over the frequency range 13.7 to 14.7 year$^{-1}$ \cite{sch02}.

The above similarity between the GSI oscillation frequencies and those found in BNL and PTB data further supports our proposal that nuclear decay rates are influenced by solar radiation that is somehow influenced by processes in the deep solar interior.  This leads us to look for a form of radiation that has two (apparently contradictory) properties: (a) the radiation can traverse the solar radiative and convection zones (and presumably the Earth) without being absorbed, and (b) the radiation can be modulated by processes in the deep solar interior and retain that modulation in the course of its passage through the solar interior.

These two requirements can be met by neutrinos: the cross section for scattering of neutrinos by normal matter is estimated to be extremely small, so that one would expect there to be negligible absorption of  the neutrino flux in its propagation from the core to Earth. However, there are believed to be at least three types of neutrinos involving the three flavors electron, muon, and tau. An inhomogeneous magnetic field (and magnetic fields are necessarily inhomogeneous) can lead to flavor oscillations by the RSFP (Resonant Spin Flavor Precession) process \cite{lim88,akh88a,akh88b}. One can imagine that such a modulation might be imprinted on the neutrino flux in the deep solar interior and remain imprinted in the flux during its passage through the outer layers of the Sun. 

The obvious and crucial objection to this proposal is that there is no known mechanism by which neutrinos can influence beta decays. Further experiments and future theoretical work will hopefully lead to a resolution of this paradox.

\section*{Acknowledgments}

The work of EF was supported in part by U.S. DOE contract No. DE-AC02-76ER071428.


\appendix

\section{Experimental Setup \cite{ste11}}
\label{app:Exp}
The experimental setup (Figure 21), located in the yard of the Geological Survey of Israel (GSI) Laboratory in Jerusalem, is contained in a square tank comprised of welded 3-mm-thick iron plates. The tank is sealed from the outer environment. The upper part of the tank contains air and the lower part contains ground phosphorite. Uranium ($^{238}$U) in the phosphorite produces radium ($^{226}$Ra) that in turn produces radon ($^{222}$Rn). A gamma-ray sensor (gamma-C), located in the upper part of the tank, is contained in a lead pipe (length 30 cm, outer diameter 10.5 cm., thickness 0.5 cm), which has a perforated lead plate at its lower end. This plate serves to reduce the direct gamma radiation from the phosphorite, while allowing gas to enter the gamma-detector chamber. The pressure and temperature sensors, a datalogger, and a 12-volt battery are located in a box under the shack. The gamma detector is a 2x2 NaI detector (PM-11 detector, ROTEM Inc., Israel), which records gamma-ray impulses of energy above 50 keV. Measurements are recorded at 15-minute intervals.

\begin{figure}[h]
\includegraphics[width=\columnwidth]{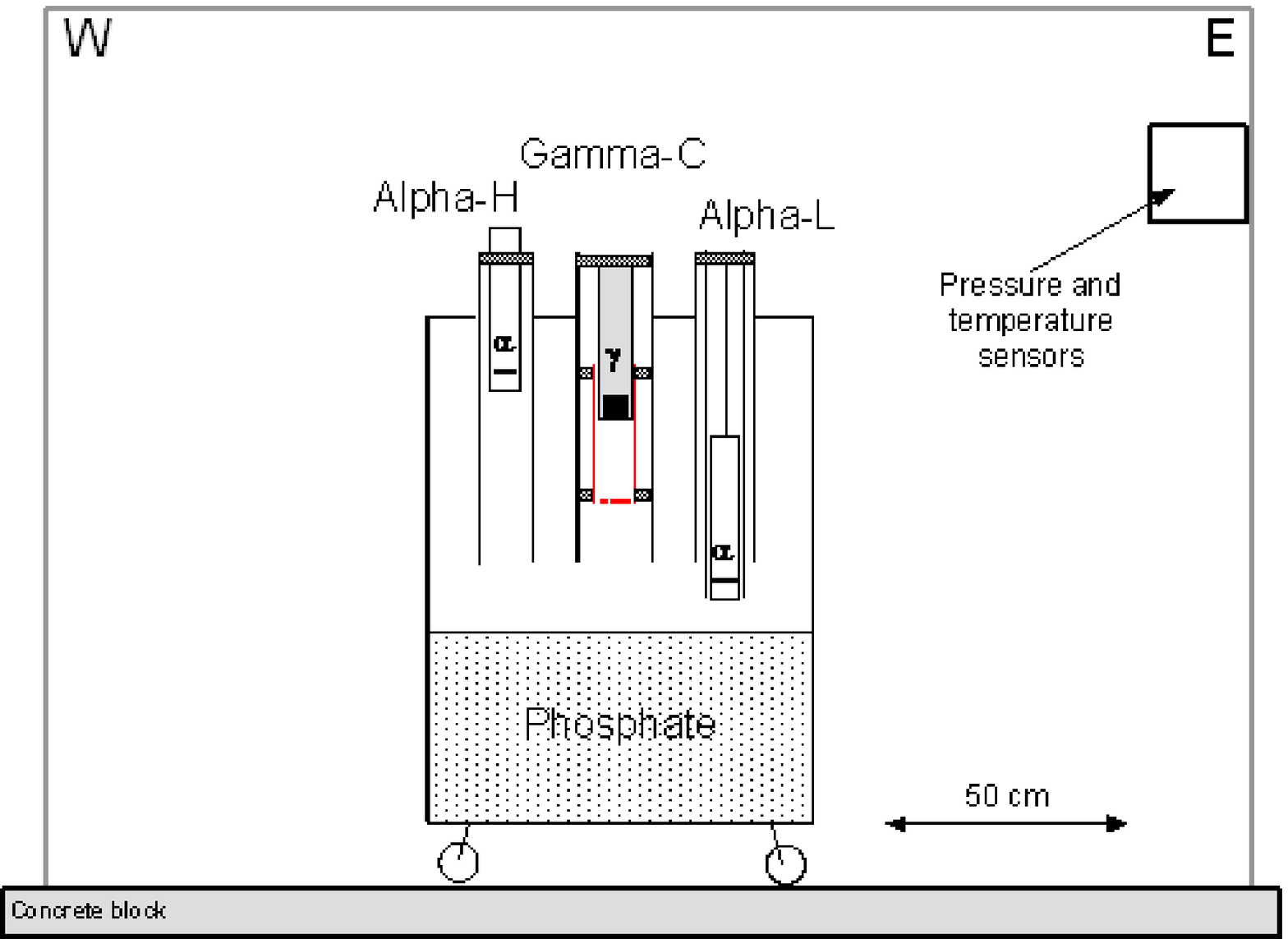}
\caption{Experimental layout, showing the locations in the tank of the phosphorite (the source of $^{222}$Rn) and the radiation detectors. This article analyzes measurements made with the gamma-C detector, which is housed located in a lead tube. Pressure and temperature sensors, a data logger, and a continuously charged 12 V battery are located in a separate enclosure attached to the wall of the shack.  \label{fig:fig21}}
\end{figure} 

\section{Environmental Drivers of Radon Production}
\label{app:Radon}
The variation patterns of radon in geogas (subsurface air) have been observed and investigated since the second half of the last century. Numerous publications aim to clarify possible reasons for observed radiation patterns and their potential geodynamical significance. Due to its noble-gas character, its constant generation, and its radioactive decay, radon provides a unique indicator of subtle variable processes in the natural environment. It is significant for geological processes that radon often shows large and complex variations that have no obvious atmospheric causes. These variations are believed to reflect subsurface processes. However, there is as yet no comprehensive theoretical understanding of observed radiation patterns. This lack of understanding of the causes of radon variability hampers its utilization as a proxy for active geodynamics. For this reason, it is important to try to understand radon variability that is not obviously related to atmospheric variability.
%% References


%\bibliographystyle{model1a-num-names}
%\bibliography{ms}






% \end{thebibliography}

\end{document}